\documentclass{amsart}
\input cyracc.def

\usepackage{amssymb}
\usepackage{amsthm}
\usepackage{amscd}
\usepackage{mathdots}

\theoremstyle{plain}
\newtheorem{thm}{Theorem}[section]
\newtheorem{lem}[thm]{Lemma}
\newtheorem{cor}[thm]{Corollary}
\newtheorem{prop}[thm]{Proposition}
\newtheorem{conj}[thm]{Conjecture}

\theoremstyle{definition}
\newtheorem{dfn}[thm]{Definition}
\newtheorem{Q}[thm]{Question}
\newtheorem{prob}[thm]{Problem}

\newtheorem{ex}[thm]{Example}

\def\dnfo{\;\raise.2em\hbox{$\mathrel|\kern-.9em\lower.4em\hbox
{$\smile$}$}}

\def\dnf#1{\lower.9em\hbox{$\buildrel\dnfo\over{ \scriptstyle  #1}$}}

\def\dfo{\;\raise.2em\hbox{$\mathrel|\kern-.9em\lower.4em\hbox{$\smile$}
\kern-.72em\lower.07em\hbox{\char'57}$}\;}
\def\df#1{\lower1em\hbox{$\buildrel\dfo\over{\scriptstyle #1}$}}

\newcommand{\fN}{\mathfrak{N}}

\newcommand{\fU}{\mathfrak{U}}

\newcommand{\sA}{\mathcal{A}}
\newcommand{\sB}{\mathcal{B}}
\newcommand{\sC}{\mathcal{C}}

\newcommand{\sG}{\mathcal{G}}

\newcommand{\sL}{\mathcal{L}}
\newcommand{\sM}{\mathcal{M}}
\newcommand{\sN}{\mathcal{N}}

\newcommand{\sP}{\mathcal{P}}

\newcommand{\sS}{\mathcal{S}}

\newcommand{\real}{\mathbb{R}}

\newcommand{\rat}{\mathbb{Q}}
\newcommand{\Z}{\mathbb{Z}}
\newcommand{\R}{\real}

\newcommand{\rg}[1]{\text{range}(#1)}

\newcommand{\bigudot}[1]{\bigcup\limits_{#1}\hspace{-0.15in}\cdot\hspace{0.14in}}

\title[Some Results on $\real$-Computable Structures]{Some Results on $\real$-Computable Structures}

\author{W.\ Calvert and J.\ E.\ Porter}
\address{Department of Mathematics \& Statistics\\ Faculty Hall 6C\\
  Murray State University\\ Murray, Kentucky 42071}
\email{wesley.calvert@murraystate.edu}

\thanks{The first author is grateful for the support of Grant \#13397
  from the Templeton Foundation.}

\begin{document}

\maketitle

\tableofcontents

\section{Introduction}

The theory of effectiveness properties on countable structures whose
atomic diagrams are Turing computable is well-studied (see, for
instance, \cite{akbook, harizanovhrm}).  Typical results describe
which structures in various classes are computable (or have
isomorphic copies that are) \cite{khishrm}, or the potential degree
of unsolvability of various definable subsets of the structure
\cite{harizanovdegspec}.  The goal of the present paper is to
survey some initial results investigating similar concerns on structures which are
effective in a different sense.

A rather severe limitation of the Turing model of computability is
its traditional restriction to the countable.  Of course, many
successful generalizations have been made (see, for instance,
\cite{sackshrt, grt, grt2, rmiller, morozovkorovina, fparam} and the
other papers in the present volume).  The
generalization that will be treated here is based on the observation
that while there is obviously no Turing machine for addition and
multiplication of real numbers, there is strong intuition that these
operations are ``computable.''  The BSS model of computation, first
introduced in \cite{bss}, approximately takes this to be the
definition of computation on a given ring (a more formal definition
is forthcoming).  This allows several problems of computation in
numerical analysis and continuous geometry to be treated rigorously.
The monograph \cite{bcss} gives the examples of the ``decision
problem'' of the points for which Newton's method will converge to a
root, and determining whether a given point is in the Mandelbrot
set.

\subsection{Basic Definitions}

The definition of a BSS machine comes from \cite{bcss}.  Such a
machine should be thought of as the analogue of a Turing machine
(indeed, the two notions coincide where $R = \mathbb{Z}$).  Let $R$
be a ring with 1.  Let $R^\infty$ be the set of finite sequences of
elements from $R$, and $R_{\infty}$ the bi-infinite direct sum
\[\bigoplus\limits_{i \in \Z} R.\]

\begin{dfn} A machine $M$ over $R$ is a finite connected directed
  graph, containing five types of nodes: input, computation, branch,
  shift, and output, with the following properties:
\begin{enumerate}
\item The unique input node has no incoming edges and only one
  outgoing edge.
\item Each computation and shift node has exactly one output edge and
  possibly several input branches.
\item Each output node has no output edges and possibly several input
  edges.
\item Each branch node $\eta$ has exactly two output edges (labeled $0_\eta$ and $1_\eta$) and possibly
  several input edges.
\item Associated with the input node is a linear map $g_I:R^\infty \to
  R_\infty$.
\item Associated with each computation node $\eta$ is a rational
  function $g_\eta: R_\infty \to R_\infty$.
\item Associated with each branch node $\eta$ is a polynomial function
  $h_\eta : R_\infty \to R$.
\item Associated with each shift node is a map $\sigma_\eta \in
  \{\sigma_l, \sigma_r\}$, where $\sigma_l(x)_i = x_{i+1}$
  and $\sigma_r(x)_i = x_{i-1}$.
\item Associated with each output node $\eta$ is a linear map $O_\eta:
  R_\infty \to R^{\infty}$.
\end{enumerate}
\end{dfn}

A machine may be understood to compute a function in the following
way:
\begin{dfn} Let $M$ be a machine over $R$.
\begin{enumerate}
\item A \emph{path} through $M$ is a sequence of nodes $(\eta_i)_{i
  =0}^n$ where $\eta_0$ is the input node, $\eta_n$ is an output node,
  and for each $i$, we have an edge from $\eta_i$ to $\eta_{i+1}$.
\item A \emph{computation} on $M$ is a sequence of pairs
  $\left((\eta_i, x_i)\right)_{i = 0}^n$ with a number $x_{n+1}$, where $(\eta_i)_{i
  =0}^n$ is a path through $M$, where $x_0 \in R^\infty$, and where,
  for each $i$, the following hold:
\begin{enumerate}
\item If $\eta_i$ is an input node, $x_{i+1} = g_I(x_i)$.
\item If $\eta_i$ is a computation node, $x_{i+1} = g_{\eta_i}(x_i)$.
\item If $\eta_i$ is a branch node, $x_{i+1} = x_i$ and $\eta_{i+1}$
  determined by $h_{\eta_i}$ so that if $h_{\eta_i}(x_i) \geq 0$, then
  $\eta_{i+1}$ is connected to $\eta_i$ by $1_{\eta_i}$ and if $h_{\eta_i}(x_i) < 0$, then
  $\eta_{i+1}$ is connected to $\eta_i$ by $0_{\eta_i}$.  (Note that
  in all other cases, $\eta_{i+1}$ is uniquely determined by the
  definition of path.)
\item If $\eta_i$ is a shift node, $x_{i+1} = \sigma_{\eta_i}(x_i)$
\item If $\eta_i$ is an output node, $x_{i+1} = O_{\eta_i}(x_i)$.
\end{enumerate}
\end{enumerate}
\end{dfn}

The proof of the following lemma is an obvious from the definitions.

\begin{lem} Given a machine $M$ and an element $z \in R^{\infty}$,
  there is at most one computation on $M$ with $x_0 = z$. \end{lem}

\begin{dfn} The function $\varphi_M : R^\infty \to R^\infty$ is
  defined in the following way:  For each $z \in R^\infty$, let
  $\varphi_M(z)$ be $x_{n+1}$, where $\left(\left((\eta_i,
  x_i)\right)_{i = 0}^n, x_{n+1}\right)$ is the unique computation, if
  any, where $x_0 = z$.  If there is no such computation, then
  $\varphi_M$ is undefined on $z$.\end{dfn}

Since a machine is a finite object, involving finitely many real
numbers as parameters, it may be coded by a member of $R^\infty$.

\begin{dfn}If $\sigma$ is a code for $M$, we define $\varphi_\sigma =
  \varphi_M$.\end{dfn}

We can now say that a set is computable if and only if its
characteristic function is $\varphi_M$ for some $M$.

\begin{ex} Let $R = \Z$.  Now the $R$-computable functions are
  exactly the classical Turing-computable functions.\end{ex}

\begin{ex} Let $R = \real$.  Then the Mandelbrot set is not
  $R$-computable (see Chapter 2 of \cite{bcss}). \end{ex}

\begin{dfn} A \emph{machine over $R$ with oracle $X$} is exactly like
  a machine over $R$, except that it has an additional type of nodes,
  the oracle nodes.  Each oracle node is exactly like a computation
  node, except that $g_{\eta} = \chi_X$.  Computations in oracle
  machines are defined in the obvious way.\end{dfn}

We say that a set $S$ is decidable (respectively, $X$-decidable) over $R$ if
and only if $S$ is both the halting set of an $R$-machine
(respectively, with oracle $X$) and the complement of the halting set
of an $R$-machine (respectively, with oracle $X$).  We also say that
$S$ is semi-decidable if and only if $S$ is the domain of an
$R$-computable function (if $R$ is a real closed field, it is
equivalent to say that $S$ is the range of an $R$-computable function \cite{bcss}).  Ziegler
\cite{zieglerrhc} gives a specialized but recent survey of results on
$\real$-computation.  The following result, first presented by Michaux,
but proved in detail in \cite{cuckersat}, is useful in characterizing
the decidable and semi-decidable sets:

\begin{prop} Let $S \subseteq \real^\infty$.  Then $S$ is
  semi-decidable if and only if $S$ is the union of a countable family
  of semialgebraic sets defined over a single finitely generated
  extension of $\rat$.\end{prop}

The ``only if'' part of this statement is the upshot of an earlier
theorem described in \cite{bcss}, called the Path Decomposition
Theorem.  We can now proceed to define computable structures.

\begin{dfn} Let $\sL = (\{P_i\}_{i \in I_P}, \{f_i\}_{i \in I_f},
    \{c_i\}_{i \in I_C})$ be a language with relation symbols
    $\{P_i\}_{i \in I_P}$, function symbols $\{f_i\}_{i \in I_f}$, and
    constant symbols $\{c_i\}_{i \in I_C}$.  Let $\sA$ be an
    $\sL$-structure with universe $A \subseteq R^\infty$.
\begin{enumerate}
\item We say that $\sL$ is $R$-computable if the sets of relations,
  functions, and constants are each decidable over $R$,
  and if, in addition, there are $R$-machines which will tell, given
  $P_i$ (respectively, $f_i$), the arity of $P_i$ (respectively,
  $f_i$).
\item We identify $\sA$ with its atomic diagram; in particular,
\item We say that $\sA$ is computable if and only if the atomic
  diagram of $\sA$ is decidable.
\end{enumerate}
\end{dfn}

The obstructions to a direct parallel between the theory of
$\real$-comptuable structures and that of Turing computable structures
which we have encountered so far are two in number (one for the
parsimonious):

\begin{enumerate}
\item The real numbers do not admit an $\omega$-like well-ordering to
  facilitate searching or priority constructions, and in particular
\item There exist $\real$-computable injective functions whose inverses are
  not $\real$-computable.
\end{enumerate}

\subsection{Plan of the Paper}

In the present paper, we will survey recent work on the theory of
$\real$-computable structures.  In Section \ref{secbasic}, we give
some basic calculations, showing some parallels with the classical
theory, including computable ordinals (Section \ref{subsecrco}),
satisfaction of computable infinitary formulas (Section
\ref{subsecsat}), and the use of forcing to carry out a simple
priority construction (Section \ref{subsecforcing}).  In Section
\ref{seccat}, we explore effective categoricity, using vector spaces
as an example.  In Section \ref{secgt}, we describe some recent
results in effective geometry and topology from the perspective of
$\real$-computation.  In Section \ref{secrelmodels} we address the
relationship of $\real$-computation with other models of effective
mathematics for uncountable structures.  In Section \ref{secconcl} we
summarize the state of $\real$-computable model theory and describe
some directions for future research.

\section{Basic Results}\label{secbasic}

\subsection{$\real$-computable Ordinals}\label{subsecrco}

The Turing computable ordinals constitute a proper initial segment of
the countable ordinals \cite{spectorrwo, kleeneord}.  This initial
segment includes, for instance, the ordinal
$\omega^{\omega^{\omega^{\iddots}}}$.  In the present section, we will
establish the following theorem:

\begin{thm} A well-ordering $(L, <)$ has an isomorphic copy which is
  $\real$-computable if and only if $L$ is countable. \end{thm}

\begin{prop} Every countable well-ordering $(M, \prec)$ has an
  isomorphic copy $(L, <)$ which is $\real$-computable. \end{prop}

\begin{proof} Since $(M, \prec)$ is countable, it has an isomorphic
  copy with universe $\omega$.  Now $D(M) = \{(a, b) \in M^2 | a \prec
  b\}$ is a subset of $\omega^2$.  Now we define a real number $\ell$
  in the following way:
\[ \ell = \sum\limits_{i \in \omega} 10^{-i} \chi_{D(M)}(i).\]
There is a $\real$ machine which, given a pair $(a,b) \in
\omega^2$ will return the $10^{-\langle a,b \rangle}$ place of $\ell$
if that place is 1 and will diverge if that place is 0.  This shows
that $D(M)$ is the halting set of a $\real$-computable function, as
required.\end{proof}

\begin{prop}\label{rco} Suppose $(L, <)$ is a $\real$-computable well-ordering.
  Then $|L| \leq \aleph_0$.\end{prop}

\begin{proof} Since $(L, <)$ is $\real$-computable, the set $L_< := \{(a,b)
  \subseteq L^2 : a < b\}$ is the halting set of a $\real$-machine.
  By Path Decomposition, it must be a disjoint union of
  semialgebraic sets, and consequently Borel.  By the Kunen-Martin Theorem (Theorem
  31.5 of \cite{kechris}), analytic (and hence Borel) well-orderings
  are countable.\end{proof}

A rather different proof of Proposition \ref{rco}, using Fubini's
Theorem, is possible and enlightening.

\begin{proof}
Since $L_<$ is uncountable and Borel, $|L_<| = 2^{\aleph_0}$.  This
implies $L$ is Borel with $|L| = 2^{\aleph_0}$.  Without loss of
generality, we suppose that $L$ is order isomorphic to the cardinal
$2^{\aleph_o}$; otherwise, an initial segment of $L$ which was
isomorphic to this cardinal would also be $\real$-computable.  In particular, $L$
contains a Cantor set $C$.  Fix a Borel measure $\mu $ on C such that
$\mu(C) = 1$ and extend $\mu $ to $L$ by setting $\mu (L\setminus
C)=0$.

We define two auxiliary sets:
\begin{eqnarray*}
L_x & = & \{b\in L : (x,b) \in L_<\}\\
L^y & = & \{a\in L : (a,y) \in L_<\}\\
\end{eqnarray*}
Each of these is a Borel set.  For any $y$, we have $|L^y| <
2^{\aleph_0}$, since $2^{\aleph_0}$ is a cardinal and $L^y$ is
isomorphic to an ordinal less than $2^{\aleph_0}$.  Since $L^y$ is
Borel, we have $|L^y| = \aleph_0$. This implies the set $L_y$ is
co-countable for any $y$.

Since $L_<$ is Borel, we can apply Fubini's theorem to calculate
$\int_{L_<} 1 d\lambda$, where $\lambda$ is the product measure $\mu
\times \mu$. On the one hand,
\[\int_{L_<} 1 d\lambda = \int_L \int_{L_x} 1 (d\mu)(d\mu) = \int _L\mu (L_x)d\mu=\int_L 1 d\mu= \mu(L)=1\]
since $L_x$ is co-countable for each $x$, and thus of full measure.
On the other hand, since $L^y$ is countable for each $y$, we have
$\int_{L_y} 1 dy = 0$.  Hence \[\int_{L_<} 1 d\lambda=\int_L
\int_{L^y} 1 (d\mu)(d\mu) =\int_L \mu(L^y) d\mu = \int_L 0d\mu =
0,\] which is a contradiction.
\end{proof}

\subsection{The Complexity of Satisfaction}\label{subsecsat}

We define the class of $\real$-computable infinitary formulas.  The definition is by analogy with the (Turing)
computable infinitary formulas already in broad usage, described in
\cite{akbook}. The choice of computable infinitary formulas is nontrivial, since
there are uncountably many $\real$-machines.  One natural approach,
not pursued here, would be to work in the $\real$-computable fragment
of $\mathcal{L}_{(2^{\aleph_0})^+, \omega}$.  This would certainly be an
interesting logic to understand, but the present authors found it more
desirable at first to understand the more familiar $\real$-computable
fragment of $\mathcal{L}_{\omega_1 \omega}$.  At issue is which
conjunctions and disjunctions are allowed in a ``computable'' formula.
The logic $\mathcal{L}_{\omega_1 \omega}$ allows countable
conjunctions and disjunctions, while $\mathcal{L}_{(2^{\aleph_0})^+,
  \omega}$ allows any of size at most $2^{\aleph_0}$.  However, the
difficulty of describing what is meant by, for instance, an interval
of formulas is a motivation (beyond the avoidance of set-theoretic
independence) to consider first the countably long formulas.

\begin{dfn} Let $\mathcal{L}$ be an $R$-computable language.
\begin{enumerate}
\item The $\Sigma_0$ formulas of $\mathcal{L}$ are exactly the
  finitary quantifier-free formulas.  The $\Pi_0$ formulas are the
  same.
\item For any ordinal $\alpha = \beta +1$, the $\Sigma^0_\alpha$
  formulas are those of the form \[\bigvee_{i \in
  S}\hspace{-.15in}\bigvee \exists \bar{y}
  [\varphi_i(\bar{x}\bar{y})]\] where $S$ is countable and is
  the halting set of an $R$-machine, and there is a finitely generated field $F\subset \mathbb{Q}$ such that
  all parameters in $\phi_i$ are in $F$.
\item For any ordinal $\alpha = \beta +1$, the $\Pi^0_\alpha$
  formulas are those of the form \[\bigwedge_{i \in
  S}\hspace{-.15in}\bigwedge \forall \bar{y}
  [\varphi_i(\bar{x}\bar{y})]\] where $S$ is countable and is
  the halting set of an $R$-machine, and there is a finitely generated field $F\subset \mathbb{Q}$ such that
  all parameters in $\phi_i$ are in $F$.
\item Suppose $\alpha = \lim\limits_n \beta_n$ where $\beta_n$ is a bounded $R$-computable sequence of ordinals,
 and there is a finitely generated field $F\subset \mathbb{Q}$ such that all parameters in $\phi_i$ are in $F$.
\begin{enumerate}
\item The $\Sigma_\alpha$ formulas are those of the form \[\bigvee_{n
  \in S}\hspace{-.15in} \bigvee \varphi_n,\] where for each $n$ the
  formula $\varphi_n$ is a $\Sigma_{\beta_n}$ formula and $S$ is countable and is
  the halting set of an $R$-machine.
\item The $\Pi_\alpha$ formulas are those of the form \[\bigwedge_{n
  \in S}\hspace{-.15in} \bigwedge \varphi_n,\] where for each $n$
  the formula $\varphi_n$ is a $\Pi_{\beta_n}$ formula and $S$ is countable and is
  the halting set of an $R$-machine.
\end{enumerate}
\end{enumerate}
\end{dfn}

The $\real$-computable infinitary formulas will be exactly the
formulas which belong to either $\Sigma_\alpha$ or
$\Pi_\alpha$ for some countable (i.e.\ $\real$-computable) $\alpha$.  Ash showed that Turing computable $\Sigma_\alpha$ formulas defined
sets which were $\Sigma^0_\alpha$ \cite{akbook}. 

We will say that a set is \emph{semantically $\real$-$\Sigma_\alpha$} if and
only if it is the set of solutions to an $\real$-computable
$\Sigma_\alpha$ formula, and similarly for $\Pi_\alpha$.  We will say
that a set is \emph{topologically $\Sigma^0_\alpha$} if it is of that
level in the standard Borel hierarchy using the order topology on $\real$.

\begin{thm} We characterize the topological structure of sets in the
  semantic hierarchy:
\begin{enumerate}
\item The semantically $\real$-$\Sigma_0$ sets are topologically
  $\Delta^0_2$.
\item If $0 < \alpha < \omega$, then the semantically
  $\real$-$\Sigma_\alpha$ sets are included among the topologically
  $\Sigma^0_{\alpha+1}$ sets.
\item If $\alpha \geq \omega$, then the semantically 
$\real$-$\Sigma_\alpha$ sets are included among the topologically
  $\Sigma^0_{\alpha}$ sets.
\end{enumerate}
\end{thm}

\begin{proof} Since $\mathcal{A}$ is a $\real$-computable structure, the
semantically $\real$-$\Sigma_0$ sets are all countable unions of
semialgebraic sets, and the completes of semantically
$\real$-$\Sigma_0$ are all countable unions of semialgebraic sets.
 Since all semialgebraic sets are topologically $\Delta^0_2$ (that is,
 both topologically $\Sigma^0_2$ and $\Pi^0_2$), the countable unions of them are all topologically $\Sigma^0_2$.
 Now if the statement holds for $n \leq k$, it clearly holds for $n = k+1$ by
  the definitions of the various classes involved. 

Toward the final statement, notice that the semantically
$\real$-$\Sigma_\omega$ sets are countable unions of sets at lower
levels, and are all topologically $\Sigma_\omega$.  Above that level,
the induction follows exactly as before.
\end{proof}

At the finite levels, Cucker proved \cite{cuckersat} that the union of
all the semantically $\real$-$\Sigma_n$ for $n < \omega$ is the class
of Borel sets of finite order.  Cucker \cite{cuckersat} defined another arithmetical hierarchy: we call a set
\emph{computationally $\Sigma_{\alpha + 1}$} if it can be enumerated by a
real machine with a computationally $\Sigma_{\alpha}$ oracle.  In
particular, the semi-decidable sets are the computationally $\Sigma_1$
sets.  Cucker proved that for all $k < \omega$, the computationally $\Sigma_k$
sets are exactly the semantically $\real$-$\Sigma_k$ sets.  It seems
likely that this result could be generalized for transfinite $\alpha$,
but we do not have a proof of this.

\subsection{Forcing as a Construction Technique}\label{subsecforcing}

Aside from the lack of inverse functions, the most difficult part of
classical computability theory to get by without is the priority
construction.  Unless this niche can be filled, we are not optimistic
concerning the parallel between Turing-computable structures and
$\real$-computable structures.  Consequently, although there are ad-hoc methods
to construct $\real$-incomparable sets \cite{meerziegler}, we give an
example in this section that is potentially more easily generalized.

\begin{prop} There exist sets $A_o$ and $A_1$ such that neither is
  computable by a $\real$-machine using the other as an
  oracle.\end{prop}

\begin{proof} The proof will closely follow the
  second proof given for the classical case by Lerman
  \cite{lermanbk}.  Let $(F,\leq)$ be the set of pairs of partial functions from
  $\real$ to $2$ whose complement contains an interval, partially ordered by extension in the sense that
  $(p_0, p_1)\leq (q_0, q_1)$ if and only if $p_i$ extends $q_i$ for
  each $i$.  It suffices to satisfy the following requirements for every $e \in \real^\infty$:
\[P_{e,i} : M_e^{A_i} \neq A_{1-i}.\]
We say that $(p_0, p_1) \Vdash P_{e,i}$ if there is some $x$ such that
either $M_e^{p_i}(x)\downarrow \neq p_{1-i}(x)$ and the latter is
defined, or for any $p$ extending $P_i$, we have $M_e^p\uparrow$.

\begin{lem}[Density Lemma] For any pair $(e,i)$, the set $\{p \in F :
  p \Vdash P_{e,i}\}$ is dense.\end{lem}

\begin{proof}
Let $q = (q_0, q_1) \in F$.  We will show that there is some $p \leq
q$ such that $p \Vdash P_{e,i}$.  Let $x$ be outside the domain of
$q_{1-i}$.  If there is no pair $s, r$ such that $r$ extends $q_i$ and
$M_{e,s}^{r}(x)
  \downarrow$, then $q \Vdash P_{e,i}$, so assume that such an $s$
  exists.  Now we extend $q_{1-i}$ by setting $p_{1-i} = q_{1-i} \cup
  \{(x, 1-M_e^{q_i}(x))\}$, and set $p_i = q_i$.
\end{proof}

The following Lemma is the only part of the construction which becomes
genuinely more difficult in the uncountable case.  

\begin{lem}[Existence of a Generic] Let $\sC$ be the collection of all
  sets of the form $\{p \in F : p \Vdash P_{e,i}\}$.  There exists a
  $\sC$-generic set; that is, a pair of functions $G = (G_0, G_1)$
  where $G_i : \real \to 2$ such that for
  each pair $(e,i)$, the function $G$ extends some element of $F$
  which forces $P_{e,i}$.\end{lem}

\begin{proof}
Let $G^0:=(\emptyset, \emptyset)$, and well-order the requirements.
We define $G^{\alpha + 1}$ to be the extension of $G^{\alpha}$ which
forces the $\alpha$th requirement.  For limit ordinals $\gamma$, we
define $G^\gamma := \bigcup\limits_{\beta < \gamma} G^\beta$.  The
union of all the $G^\alpha$ is a $\sC$-generic.
\end{proof}

Of course, we may take $G$ to be total, by setting all undefined
values to $0$.  Now we take $A_0$ to be the set whose characteristic
function is $G_0$ and $A_1$ the set with characteristic function $G_1$.
\end{proof}

\section{Effective Categoricity}\label{seccat}

It is often possible to produce two classically computable structures
which are isomorphic, but for which the isomorphism is not witnessed
by a computable function.  Any theory of effective mathematics must
take account of this phenomenon.  

\begin{dfn}\label{ccat} A computable structure $\sM$ is said to be
  \emph{computably categorical} if and only if for any computable
  structure $\sN \simeq \sM$ there is a computable function $f:\sN
  \stackrel{\simeq}{\to} \sM$.  The number of equivalence classes
  under computable isomorphism contained in an isomorphism type is
  called its \emph{computable dimension}.\end{dfn}

In the present section, we describe progress toward a parallel to the
following classical result:

\begin{thm}[see \cite{nurtazin}, although it was almost certainly known earlier]\label{classvs} If $V$ is a countable vector space over $\rat$,
\begin{enumerate}
\item There is a Turing-computable copy of $V$,
\item The categoricity properties are as follows:
\begin{enumerate}
\item If $dim(V)$ is finite, then $V$ is computably categorical, and
\item If $dim(V) = \omega$, then the computable dimension of $V$ is
  $\omega$.
\end{enumerate}
\end{enumerate}
\end{thm}

The existence part of the theorem is still true without serious modification.

\begin{prop} Let $n \in \aleph_0 \cup \{\aleph_0, 2^{\aleph_0}\}$.
  Then there is a $\real$-computable vector space $V^n$ of dimension
  $n$.  Further, $V^n$ has a $\real$-computable basis.\end{prop}

\begin{proof} Consider the language of real vector spaces (addition,
  plus one scaling operation for each element of $\real$).  Let $\{b_i
  : i \in I\}$ be a $\real$-computable set of constants, where $|I| = n$.  The set of
  closed terms with constants from $\{b_i: i \in I\}$, modulo provable
  equivalence (in the theory of vector spaces) is a model of the
  theory of vector spaces, and has dimension $n$.\end{proof}

Of course, the categoricity result highlights an additional concern
with $\real$-comp\-ut\-at\-ion: It may happen that there is a
$\real$-computable isomorphism with no $\real$-computable inverse.
Thus, while the following result establishes, according to Definition
\ref{ccat}, something very close to part 2a of Theorem \ref{classvs},
it falls short of full analogy.

\begin{prop} Let $n < \aleph_0$.  Then for any real vector space $W$
  of dimension $n$, there is a computable isomorphism $f:V^n \to W$.\end{prop} 

\begin{proof}
Let $\{a_1, \dots, a_n\}$ be a basis of $W$.  Each member of $V^n$ is a
  $\real$-linear combination $\sum\limits_{i=1}^n \lambda_i b_i$.  We
  map $\sum\limits_{i=1}^n \lambda_i b_i$ to $\sum\limits_{i=1}^n \lambda_i a_i$.
\end{proof}

The classical way to prove part 2b of Theorem \ref{classvs} is to produce a
computable vector space with a computable basis, and an isomorphic
(i.e.\ same dimension) vector space with no computable basis.  Without
recourse to priority constructions, this strategy seems, for the
present, very difficult in the $\real$-computable context.

\section{Geometry and Topology}\label{secgt}

In the talk by the first author at EMU 2008, an early slide asked for
a context in which one could formulate effectiveness questions for
results like Thom's Theorem on cobordism or the classification of
compact 2-manifolds.  Some work in the intervening months, which began
at that meeting, has yielded interesting results in $\real$-computable
topology.

An $n$-manifold is a topological space which is locally homeomorphic
to $\real^n$, satisfying some fairly obvious regularity conditions on
the intersections of the neighborhoods on which homeomorphism holds.
The following definition is given in \cite{calvertmillerpi1}.

\begin{dfn}
\label{defn:manifold} 
A \emph{real-computable $d$-manifold} $M$ consists of
real-computable $i$, $j$, $j'$, $k$, the \emph{inclusion
functions}, satisfying the following conditions for all
$m,n\in\omega$.
\begin{itemize}
\item
If $i(m,n)\downarrow =1$, then $\phi_{j(m,n)}$ is a total
real-computable homeomorphism from $\R^d$ into $\R^d$,
and $\phi_{j'(m,n)}=\phi_{j(m,n)}^{-1}$,
and $k(m,n)\downarrow =k(n,m)\downarrow =m$.
\item
If $i(m,n)\downarrow =0$, then $k(m,n)\downarrow =
k(n,m)\downarrow\in\omega$ with
$i(k(m,n),m)=i(k(m,n),n)=1$ and for all $p\in\omega$, if
$i(p,m)=i(p,n)=1$, then $i(p,k(m,n))=1$, and for all $q\in\omega$,
if $i(m,q)=i(n,q)=1$, then $i(k(m,n),q)=1$ with \[\rg{\phi_{j(m,q)}}
\cap \rg{\phi_{j(n,q)}} = \rg{\phi_{j(k(m,n),q)}}.\]
\item
If $i(m,n)\notin\{ 0,1\}$, then $i(m,n)\downarrow
=i(n,m)\downarrow =-1$, and
\[(\forall p\in\omega)[i(p,m)\neq 1\text{~or~} i(p,n)\neq 1],\] and
for all $q\in\omega$, if $i(m,q)$ and $i(n,q)$ both lie in $\{ 0,1\}$,
then \[\rg{\phi_{j(k(m,q),q)}}\cap\rg{\phi_{j(k(n,q),q)}}=\emptyset.\]
\item
For all $q\in\omega$, if $i(m,n)=i(n,q)=1$, then $i(m,q)=1$ and
$$ \phi_{j(n,q)}\circ\phi_{j(m,n)}
=\phi_{j(m,q)}.$$
\end{itemize}
\end{dfn}

In essence, each natural number $m$ represents a chart $U_m$.  The
functions $i(m,n)$ tell whether $U_m$ is a subset of $U_n$ and whether
$U_n$ is a subset of $U_m$.  The function $j(m,n)$ is the index for a
computable map giving the inclusion of $U_m$ in $U_n$.

\subsection{Classifying Compact 2-Manifolds}\label{subsec2mfd}

Classification of $n$-manifolds up to homeomorphism in general is
quite difficult.  However, a well-known theory of disputed priority
offers the following classification of compact connected 2-manifolds.

\begin{thm} Let $X$ be a compact connected 2-manifold.  Then $X$ is
  homeomorphic to a connected sum of $2$-spheres, copies of
  $\mathbb{RP}^2$, copies of $S^1 \times S^1$, and copies of the Klein
  Bottle.\end{thm}

Unpublished work by the first author and Montalban, inspired in part
by discussions with R.\ Miller, gives an effective version of this
result.

\begin{thm}[Calvert--Montalban] Let $M$ and $N$ be $\real$-computable
  compact $2$-manifolds.  Then there is a $\real$-computable
  homeomorphism $f: M \to N$.\end{thm}

\begin{proof}[Proof outline.] We can triangulate each of $M$ and $N$
  to form a finite simplicial complex.  The function $f$ consists of a
  mapping on the complexes, with a smoothing effect.\end{proof}

\begin{cor} Let $X$ be a compact connected $\real$-computable
  $2$-manifold.   Then $X$ is homeomorphic by a $\real$-computable
  function to a connected sum of $2$-spheres, copies of
  $\mathbb{RP}^2$, copies of $S^1 \times S^1$, and copies of the Klein
  Bottle.\end{cor}

\subsection{Computing Homotopy Groups}\label{subsechom}

One standard set of topological invariants for a manifold $M$ is the
sequence of groups $(\pi_n(M))_{n \in \omega}$, where $\pi_n(M)$ is
the group of continuous mappings from $S^n$ to $M$, up to homotopy
equivalence.  Under the classical model of computation, manifolds are often
represented by simplicial complexes in order to discuss the
possibility of computing various topological invariants.  Brown showed
\cite{brown} that there is a procedure which will, given a finite simplicial
complex $M$, compute a set of generators and relations for each of the
groups $\pi_n(M)$.  It is natural to ask, now that we have a notion of
computation that gives us algorithmic access to the manifolds
themselves, whether this can be computed directly from the manifolds.
We restrict attention here to the case of $\pi_1$, studied in detail
in \cite{calvertmillerpi1}, although it is likely that similar results
could be established for $\pi_n$.

\begin{lem}[Calvert--Miller \cite{calvertmillerpi1}] Every loop $f$ in
  a computable manifold $M$ is homotopic to a computable loop in $M$
  whose only real parameters are the base point and the inclusion
  functions necessary to define $M$.\end{lem}

Nevertheless, the answer to the question of computing a fundamental
group from a manifold is largely negative:

\begin{thm}[Calvert--Miller \cite{calvertmillerpi1}] Let $M$ be a
  $\real$-computable manifold which is connected but not simply
  connected.  Then there is no algorithm to decide whether a given
  loop is nullhomotopic.\end{thm}

\begin{thm}[Calvert--Miller \cite{calvertmillerpi1}] There is no
  $\real$-computable function which will decide, given a
  $\real$-computable manifold, whether that manifold is simply
  connected.\end{thm}

Nevertheless, there is a canonical family of loops, sufficient to
represent the whole (but not recoverable by a uniform procedure) from
which we could make the necessary computations for a fundamental
group.

\begin{lem}[Calvert--Miller \cite{calvertmillerpi1}] Let $M$ be a
  $\real$-computable manifold.  Then there is a $\real$-computable
  function $S_M$, defined on the naturals, such that the set $S_M(n)$
  consists of a set of indices for loops and contains exactly one
  representative from each homotopy equivalence type.\end{lem}

While we cannot effectively pass from an index for $M$ to an index for
$S_M$, this step includes all of the difficulty in computing
$\pi_1(M)$:

\begin{thm}[Calvert--Miller \cite{calvertmillerpi1}] Let $M$ be a
  $\real$-computable manifold.  Then there is a uniform procedure to
  pass from an index for $S_M$ to an index for a real-computable
  presentation of the group $\pi_1(M)$.\end{thm}

\section{Relations with Other Models}\label{secrelmodels}

\subsection{Local Computability}\label{subseclocal}

Let $T$ be a $\forall$-axiomatizable theory in a language with $n$
symbols.

\begin{dfn} A \emph{simple cover} of $\mathcal{S}$ is a (finite or
countable) collection $\fU$ of finitely generated models $\sA _0,\sA
_1,...$ of $T$, such that:
\begin{itemize}
\item[-] every finitely generated substructure of $\sS$ is
isomorphic to some $\sA_i\in \fU$; and
\item[-] every $\sA_i\in \fU$ embeds isomorphically into $\sS$.
\end{itemize}

A simple cover $\fU$ is \emph{computable} if every $\sA_i\in \fU$ is
a computable structure whose domain is an initial segment of
$\omega$.  $\fU$ is \emph{uniformly computable} if the sequence
$\langle (\sA_i,\overline{a}_i)\rangle _{i\in\omega}$ can be given
uniformly: there must exist a computable function which, on input
$i$, outputs a tuple of elements $\langle e_1,...,e_n,\langle
a_0,...,a_k\rangle \rangle \in \omega^n\times \sA_i^{<\omega}$ such
that $\{a_0,...,a_{k_i}\}$ generates $\sA_i$ and $\phi_{e_j}$
computes the $j$-th function, relation, or constant in $\sA_i$.

\end{dfn}


\begin{dfn}
An embedding $f:\sA_i\hookrightarrow \sA_j$ \emph{lifts} to the
inclusion $\sB\subset \sC$, via isomorphisms $\beta :\mathcal{A}_i
\twoheadrightarrow\sB$ and $\gamma :\mathcal{A}_j
\twoheadrightarrow\sC$, if the diagram below commutes:
$$\begin{CD}
\sB @>>\subseteq >\sC\\
@A\beta A\cong A @A\gamma A\cong A\\
\sA_i @>f>> \sA_j\\
\end{CD}\hskip 40pt \text{with}\hskip 4pt \gamma \circ f = \beta$$
A \emph{cover} of $\sS$ consists of a simple cover
$\fU=\{\sA_0,\sA_1,...\}$ of $\sS$, along with sets $I_{ij}^{\fU}$
(for all $\sA_i,\sA_j\in\fU$) of injective homomorphisms
$f:\sA_i\hookrightarrow\sA_j$, such that:
\begin{enumerate}
\item for all finitely generated substructures $\sB\subseteq\sC$ of
$\sS$, there exists $i,j\in \omega$ and an $f\in I_{ij}^{\fU}$ which
lifts to $\sB\subseteq\sC$ via some isomorphisms $\beta
:\mathcal{A}_i \twoheadrightarrow\sB$ and $\gamma :\mathcal{A}_j
\twoheadrightarrow\sC$; and
\item for every $i$ and $j$, every $f\in I_{ij}^{\fU}$ lifts to an
inclusion $\sB\subseteq\sC$ in $\sS$ via some isomorphism $\beta$
and $\gamma$.
\end{enumerate}
This cover is \emph{uniformly computable} if $\fU$ is a uniformly
computable simple cover of $\sS$ and there exists a c.e. set $W$
such that for all $i,j\in\omega$
$$I_{ij}^{\fU}=\{\phi_e\upharpoonright\sA_i:\langle i,j,e\rangle \in
W\}.$$

A structure $\sB$ is \emph{locally computable} if it has a uniformly
computable cover.
\end{dfn}

\begin{prop}[\cite{rmiller}]
A structure $\sS$ is \emph{locally computable} if and only if it has
a uniformly computable simple cover.
\end{prop}

\begin{prop}[\cite{rmiller}] The ordered field of real numbers is not
  locally computable.\end{prop}

However, the ordered field of real numbers is trivially
$\real$-computable.  It appears at first that the ordering might be
essential in escaping local computability.

\begin{dfn} A $\real$-machine is said to be \emph{equational} if and
  only if each branch node is decided by a polynomial
  $\emph{equation}$.  We call a structure \emph{equationally
  $\real$-computable} if its diagram is computable by an equational
  $\real$-machine.
\end{dfn}

\begin{lem}[Path Decomposition for Equational Machines] Let $M$ be an
  equational $\real$-machine.  Then the halting set of $M$ is a
  countable disjoint union of algebraic sets.\end{lem}

\begin{proof}The proof is exactly the same as for normal
  $\real$-machines.\end{proof}

\begin{cor} The ordered field of real numbers is not equationally
  $\real$-computable.\end{cor}

\begin{thm} There is an equationally $\real$-computable structure
  which is not locally computable.\end{thm}

\begin{proof} Let $S$ be a noncomputable set of natural numbers, and
  denote by $C_n$ a cyclic graph on $n$ vertices (i.e.\ an $n$-gon).
  Now let $\sG$ be the structure given by

\[\left(\bigudot{n \in S} C_{2n} \right)
\bigcup \hspace{-0.11in}\cdot\hspace{0.09in} \left(\bigudot{n \notin S} C_{2n+1} \right).\]

To show that $\sG$ is equationally $\real$-computable, we observe
that the disjoint union of two $\real$-computable structures is
$\real$-computable (since the same is true of the cardinal sum).  However, each of the graphs $C_k$ has a
$\real$-computable copy by Lagrange interpolation.

Suppose $f$ is a uniform computable enumeration of the finitely
generated substructures of $\sG$.  Then we could compute whether $n
\in S$ by searching the structures indexed by $f(t)$ for successive
$t$ until we see a substructure of type $C_{2n}$ or of type
$C_{2n+1}$.  Since $S$ is noncomputable, no such $f$ can exist, so
that $\sG$ is not locally computable.
\end{proof}

\begin{thm}\label{lcstruct} There is a locally computable structure which is not
  $\real$-computable.\end{thm}

\begin{proof} Let $X$ be the set of all countable graphs with universe
  $\omega$, and let $E$ be the isomorphism relation on $X$.  Now $E$
  is complete analytic \cite{fs}, so $F = E^c$ is complete
  co-analytic.  Now for any $x \in X$, we have $\lnot x F x$, and for
  any $x, y \in X$ we have $x F y$ if and only if $y F x$.  Thus, $F$
  defines the adjacency relation of a graph on $X$.  Let $\mathcal{X}$
  denote the graph $(X, F)$.

Now $\mathcal{X}$ is not real-computable, since its diagram is
complete co-analytic (contradicting path decomposition).  We will
show that $\mathcal{X}$ is locally computable.  Now the finitely
generated substructures of $\mathcal{X}$ are all finite graphs, and
it only remains to determine which finite graphs are included.  Let
$T$ be the following graph:

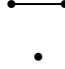
\begin{center}
\begin{picture}(40,40)
\put(25,15){\circle*{3}} \put(15,35){\circle*{3}}
\put(35,35){\circle*{3}} \put(15,35){\line(1,0){20}}
\end{picture}
\end{center}

Let $\sG$ be a finite $T$-free graph.  We will show that $\sG$ embeds in
$\mathcal{X}$.  Let $\sG = (\{0, \dots, n\}, G)$.  We will define an
equivalence structure $R$ with universe $N = \{0, \dots, n\}$.  For $x,
y \in N$, we say that $xRy$ if and only if $\lnot xGy$.  This
relation $R$ will be reflexive and symmetric.  Since $G$ is
$T$-free, $R$ will also be transitive.  Now since the isomorphism
relation is Borel complete \cite{fs}, there is a function $f:N \to
X$ such that $xRy$ if and only if $f(x) E f(y)$.  This function can
also be required to be injective \cite{hkulm}.

Now let $\Phi$ be a computable Friedberg enumeration of finite
graphs up to isomorphism (i.e.\ a total computable function whose
range consists of an index for exactly one representative from each
isomorphism class of finite graphs).  Such an enumeration was given
in \cite{cckm}.  We will define a Friedberg enumeration $\Psi$ of
finite $T$-free graphs up to isomorphism as follows: $\Psi(x)$ will
be $\Phi(x')$ for the least $x'$ such that $\Phi(x')$ is $T$-free
and $\Phi(x') \notin ran(\Psi|_x)$.  Since all of the graphs are
finite, we can effectively check whether each is $T$-free, so that
$\Psi$ is computable.  Now $\Psi$ provides a uniform simple
computable cover for $\mathcal{X}$.\end{proof}

\begin{cor} There is another structure $\tilde{\mathcal{X}}$ with the
  same uniform simple computable cover as $\mathcal{X}$, such that
  $\tilde{\mathcal{X}}$ is $\real$-computable.\end{cor}

\begin{proof} Let $\tilde{\mathcal{X}}$ be the disjoint union
  \[\bigudot{x \in \omega} \Psi(x).\]  Now
  $\tilde{\mathcal{X}}$ is countable, and so is trivially
  $\real$-computable.\end{proof}

One can say more about the structure described in Theorem
\ref{lcstruct}.  The structure satisfies a stronger condition called \emph{perfectly local computability}.  We
recall the definition of perfectly locally computable and leave the
details to the reader.

\begin{dfn}
Let $\fU$ be a uniformly computable cover for a structure $\sS$.  A
Set $M$ is a \emph{correspondence system} for $\fU$ and $\sS$ if it
satisfies all of the following:
\begin{enumerate}
\item Each element of $M$ is an embedding of some $\sA_i\in \fU$
into $\sS$; and
\item Every $\sA_i\in \fU$ is the domain of some $\beta \in M$; and
\item Every generated $\sB\subset \sS$ is the image of some
$\beta\in M$; and
\item For every $i$ and $j$ and every $\beta \in M$ with domain
$\sA_i$, every $f\in I^{\fU}_{ij}$ lifts to an inclusion $\beta
(\sA_i)\subset \gamma (\sA_i) $ via $\beta $ and some $\gamma \in
M$; and
\item For every $i$, every $\beta\in M$ with domain $\sA_i$, and
every finitely generated $\sC\subset \sS$ containing $\beta(\sA_i)$,
there exist a $j$ and an $f\in I^{\fU}_{ij}$ which lifts to $\beta
(\sA_i)\subset \sC$ via $\beta$ and some $\gamma
:\sA_j\twoheadrightarrow \sC\in M$.
\end{enumerate}
The correspondence system is \emph{perfect} if it also satisfies
\begin{enumerate}
\item[6.] For every finitely generated $\sB\subset \sS$, if
$\beta:\sA_i\twoheadrightarrow \sB$ and $\gamma :\sA_j
\twoheadrightarrow \sB$ both lie in $M$ and have image $\sB$, then
$\gamma^{-1}\circ\beta\in I_{ij}^{\fU}$.
\end{enumerate}
If a perfect correspondence system exists, then its elements are
called \emph{perfect matches} between their domains and their
images.  $\sS$ is then said to be \emph{perfectly locally
computable} with \emph{perfect cover} $\fU$.
\end{dfn}

\subsection{$\Sigma$-Definability}\label{subsecsigmadef}

The following definition is standard,
and appears in equivalent forms in \cite{barwisebk} and 
\cite{ershovopred}.

\begin{dfn} Given a structure $\sM$ with universe $M$, we define a new
  structure $HF(\sM)$ as follows.
\begin{enumerate}
\item The universe of $HF(\sM)$ is the union of the chain
  $HF_n(M)$ defined as follows:
\begin{enumerate}
\item $HF_0(M) = M$
\item $HF_{n + 1}(M) = \sP^{< \omega}\left(M \cup
  HF_n(M)\right)$, where $\sP^{< \omega}(S)$ is the set of all finite
  subsets of $S$
\end{enumerate}
\item The language for $HF(\sM)$ consists of a unary predicate $U$ for
  $HF_0(M)$, as well as a predicate $\in$
  interpreted as membership, plus a symbol $\sigma^*$ for
  each symbol $\sigma$ of the language of $\sM$, given the
  interpretation of $\sigma$ on $M = HF_0(M)$.
\end{enumerate}
\end{dfn}

Ershov gave a definition \cite{ershovopred} of a notion
generalizing computability to structures other than $\fN$.  We will
first give Barwise's definition \cite{barwisebk} of the class of $\Sigma$-formulas.

\begin{dfn} The class of $\Sigma$-formulas are defined by induction.
\begin{enumerate}
\item Each $\Delta_0$ formula is a $\Sigma$-formula.
\item If $\Phi$ and $\Psi$ are $\Sigma$-formulas, then so are
  $(\Phi \wedge \Psi)$ and $(\Phi \vee \Psi)$.
\item For each variable $x$ and each term $t$, if $\Phi$ is a
  $\Sigma$-formula, then the following are also $\Sigma$-formulas:
\begin{enumerate}
\item $\exists (x \in t) \ \  \Phi$
\item $\forall (x \in t) \ \ \Phi$, and
\item $\exists x \Phi$.
\end{enumerate}
\end{enumerate}
\end{dfn}

A predicate $S$ is called a \emph{$\Delta$-predicate} if both $S$ and
its complement are defined by $\Sigma$-formulas.

\begin{dfn} Let $\sM$ and $\sN = (N, P_0, P_1, \dots)$ be structures.
  We say that $\sN$ is $\Sigma$-definable in $HF(\sM)$ if and only if
  there are $\Sigma$-formulas $\Psi_0, \Psi_1, \Psi_1^*, \Phi_0,
  \Phi_0^*, \Phi_1, \Phi_1^*, \dots$ such that
\begin{enumerate}
\item $\Psi_0^{HF(\sM)} \subseteq HF(\sM)$ is nonempty,
\item $\Psi_1$ defines a congruence relation on $\left(\Psi_0^{HF(\sM)},
  \Phi_0^{HF(\sM)}, \Phi_1^{HF(\sM)}, \dots\right)$,
\item $(\Psi_1^{*})^{HF(\sM)}$ is the relative complement in
  $(\Psi_0^{HF(\sM)})^2$ of $\Psi_1^{HF(\sM)}$,
\item For each $i$, the set $(\Phi_i^{*})^{HF(\sM)}$ is the relative complement in
  $\Psi_0^{HF(\sM)}$ of $\Phi_i^{HF(\sM)}$, and
\item $\sN \simeq \left(\Psi_0^{HF(\sM)}, \Phi_0^{HF(\sM)},
  \Phi_1^{HF(\sM)}, \dots\right)/_{\Psi_1^{HF(\sM)}}$.
\end{enumerate}
\end{dfn}

\begin{thm}[Calvert \cite{fparamwc}]\label{sdefrc} The structures which have
  isomorphic copies $\Sigma$-definable over $HF(\real)$ are exactly
  the ones which have isomorphic copies which are
  $\real$-computable.\end{thm}

An interesting consequense of this (an immediate corollary of Theorem
\ref{sdefrc} and a result of Morozov and Korovina
\cite{morozovkorovina}) gives a sense in which some $\real$-computable
structures can be approximated by classically computable structures.

\begin{dfn} Let $\sA$ and $\sB$ be structures in a common signature.
  We write that $\sA \leq_1 \sB$ if $\sA$ is a substructure of $\sB$,
  and for all existential formulas $\varphi(\bar{x})$ and for all
  tuples $\bar{a} \subseteq \sA$, we have \[\sB \models
  (\varphi(\bar{a}) \Rightarrow \sA \models
  \varphi(\bar{a}).\]\end{dfn}

\begin{cor}\label{capprox} For any $\real$-computable structure $\sM$
  whose defining machine involves only algebraic reals as parameters,
  there is a computable structure $\sM^*$ such that $\sM^* \leq_1
  \sM$.\end{cor}

\subsection{F-Parameterizability}\label{subsecfparam}

Morozov introduced a concept that he called
\emph{$F$-par\-am\-et\-er\-iz\-a\-bil\-ity} in order to understand the elementary
substructure relation on both automorphism groups and the structure of
hereditarily finite sets over a given structure \cite{fparam}.  In a talk at
Stanford University, though, he identified this notion as one ``which
generalizes the notion of computable'' \cite{morozovstanford}.

\begin{dfn}[\cite{fparam}]\label{df:fparam} Let $\sM$ be a structure
  in a finite relational language $\left(P_n^{k_n})_{n \leq k}\right)$.  We
  say that $\sM$ is $F$-parameterizable if and only if there is an
  injection $\xi : \sM \to \omega^\omega$ with the following
  properties:
\begin{enumerate}
\item The image of $\xi$ is analytic in the Baire space, and
\item For each $n$, the set $\left\{\left(\xi(a_i)\right)_{i \leq k_n} :
  \sM \models P_n (\bar{a})\right\}$ is analytic.
\end{enumerate}
\end{dfn}

The function $\xi$ is called an $F$-parameterization of $\sM$.
Morozov also introduced the following stronger condition, essentially
requiring that $\sM$ be able to define its own
$F$-parameterization.

\begin{dfn}[\cite{fparam}]\label{fsparam} Let $\sM$ be an
  $F$-parameterizable structure.  We say that $\sM$ is weakly
  selfparameterizable if and only if there are functions $\Xi, p:
  \sM \times \omega \to \omega$, both definable without parameters in
  $HF(\sM)$, with the following properties:
\begin{enumerate}
\item For all $x \in \sM$ and all $m \in \omega$, we have $\Xi(x,m) =
  \xi(x)[m]$, and
\item For all $f \in \omega^\omega$ there is some $x \in \sM$ such
  that for all $n \in \omega$ we have $p(x,n) = f(n)$.
\end{enumerate}
\end{dfn}

In making sense of effectiveness on uncountable structures, a major
motivation is to describe a sense in which real number arithmetic ---
an operation that, while not Turing computable, does not seem horribly
ineffective --- can be considered to be effective.

\begin{prop}[Morozov \cite{fparam}]\label{realsarefparam} The real field is weakly
  $F$-selfparameterizable. \end{prop}

\begin{proof}[Outline of proof.] Define a function $\xi:\real \to
  \omega^\omega$ maps $x$ to its decimal expansion.  This function is
  definable without parameters in $HF(\real)$, in the sense
  required by Definition \ref{fsparam}.
\end{proof}

\begin{thm}[Calvert \cite{fparamwc}] Every $\real$-computable
  structure is $F$-parameterizable.  On the other hand, the structure
  $(\real, +, \cdot, 0, 1, e^x)$ is weakly $F$-selfparameterizable but
  not $\real$-computable.\end{thm}

\section{Conclusion}\label{secconcl}

We state here some open problems arising from issues discussed in the
present paper.  The first is perhaps the most vital.

\begin{prob} Develop a substitute for the priority method which is
  capable of handling constructions with injury.\end{prob}

\begin{Q} Is it true that for any $\real$-computable finite
  dimensional $\real$-vector spaces $M$ and $N$ with the same
  dimension, there is a $\real$-computable isomorphism from $M$ to
  $N$?\end{Q}

\begin{conj} A $\real$-computable $\real$-vector space of dimension
  greater than $\aleph_0$ is not $\real$-computably
  categorical.\end{conj}

We would also like to know about the categoricity of vector spaces of
dimension $\aleph_0$, but are not ready to hazard a conjecture at this
time.

\begin{Q} Does there exist a $\real$-computable Banach space of
  infinite dimension in the
  language of vector spaces, augmented by a sort for $\real$ and a
  function interpreted as the norm?\end{Q}

\begin{Q} Does there exist a $\real$-computable Hilbert space of
  infinite dimension in the
  language of vector spaces, augmented by a sort for $\real$ and a
  binary function interpreted as the inner product?\end{Q}

On each of the previous two questions, the authors had difficulty
guaranteeing completeness.

\bibliographystyle{amsplain}
\bibliography{rcmt}

\end{document}